\documentclass[twoside,twocolumn,9pt]{article}
\usepackage{extsizes}
\usepackage[super,sort&compress,comma]{natbib} 
\usepackage{multicol}
\usepackage[version=3]{mhchem}
\usepackage[left=1.5cm, right=1.5cm, top=1.785cm, bottom=2.0cm]{geometry}
\usepackage{balance}
\usepackage{gensymb}
\usepackage{mathptmx}
\usepackage{sectsty}
\usepackage{graphicx} 
\usepackage{lastpage}
\usepackage[format=plain,justification=justified,singlelinecheck=false,font={stretch=1.125,small,sf},labelfont=bf,labelsep=space]{caption}
\usepackage{float}
\usepackage{fancyhdr}
\usepackage{fnpos}
\usepackage[english]{babel}
\addto{\captionsenglish}{%
  
}
\usepackage{array}
\usepackage{droidsans}
\usepackage{charter}
\usepackage[T1]{fontenc}
\usepackage[usenames,dvipsnames]{xcolor}
\usepackage{setspace}
\usepackage[compact]{titlesec}
\usepackage{hyperref}
\usepackage{comment}

\usepackage{epstopdf}

\usepackage{upgreek}

\definecolor{cream}{RGB}{222,217,201}

\begin{document}

\pagestyle{fancy}
\thispagestyle{plain}
\fancypagestyle{plain}{
\renewcommand{\headrulewidth}{0pt}
}

\makeFNbottom
\makeatletter
\renewcommand\LARGE{\@setfontsize\LARGE{15pt}{17}}
\renewcommand\Large{\@setfontsize\Large{12pt}{14}}
\renewcommand\large{\@setfontsize\large{10pt}{12}}
\renewcommand\footnotesize{\@setfontsize\footnotesize{7pt}{10}}
\makeatother

\renewcommand{\thefootnote}{\fnsymbol{footnote}}
\renewcommand\footnoterule{\vspace*{1pt}%
\color{cream}\hrule width 3.5in height 0.4pt \color{black}\vspace*{5pt}} 
\setcounter{secnumdepth}{5}

\makeatletter 
\renewcommand\@biblabel[1]{#1}            
\renewcommand\@makefntext[1]%
{\noindent\makebox[0pt][r]{\@thefnmark\,}#1}
\makeatother 
\renewcommand{\figurename}{\small{Fig.}~}
\sectionfont{\sffamily\Large}
\subsectionfont{\normalsize}
\subsubsectionfont{\bf}
\setstretch{1.125} 
\setlength{\skip\footins}{0.8cm}
\setlength{\footnotesep}{0.25cm}
\setlength{\jot}{10pt}
\titlespacing*{\section}{0pt}{4pt}{4pt}
\titlespacing*{\subsection}{0pt}{15pt}{1pt}

\fancyfoot{}
\fancyfoot[RO]{\footnotesize{\sffamily{1--\pageref{LastPage} ~\textbar  \hspace{2pt}\thepage}}}
\fancyfoot[LE]{\footnotesize{\sffamily{\thepage~\textbar\hspace{3.45cm} 1--\pageref{LastPage}}}}
\fancyhead{}
\renewcommand{\headrulewidth}{0pt} 
\renewcommand{\footrulewidth}{0pt}
\setlength{\arrayrulewidth}{1pt}
\setlength{\columnsep}{6.5mm}
\setlength\bibsep{1pt}

\makeatletter 
\newlength{\figrulesep} 
\setlength{\figrulesep}{0.5\textfloatsep} 

\newcommand{\topfigrule}{\vspace*{-1pt}%
\noindent{\color{cream}\rule[-\figrulesep]{\columnwidth}{1.5pt}} }

\newcommand{\botfigrule}{\vspace*{-2pt}%
\noindent{\color{cream}\rule[\figrulesep]{\columnwidth}{1.5pt}} }

\newcommand{\dblfigrule}{\vspace*{-1pt}%
\noindent{\color{cream}\rule[-\figrulesep]{\textwidth}{1.5pt}} }

\makeatother

\twocolumn[
  \begin{@twocolumnfalse}
\vspace{1em}
\sffamily
\begin{tabular}{m{0.1cm} p{17.9cm} }

 & \noindent\LARGE{\textbf{Tutorial for the growth and development of \textit{Myxococcus xanthus} as a Model System at the Intersection of Biology and Physics$^\dag$}} \\
\vspace{0.3cm} & \vspace{0.3cm} \\

 & \noindent\large Jesus Manuel Antúnez Domínguez,
 \textit{$^{a}$}
 Laura Pérez García, \textit{$^{a}$}
 Natsuko Rivera-Yoshida,
 \textit{$^{b}$}
 Jasmin Di Franco,
 \textit{$^{c,d}$}
 David Steiner,
 \textit{$^{c}$}
 Alejandro V. Arzola, \textit{$^{e}$}
 Mariana Benítez, \textit{$^{b}$} 
 Charlotte Hamngren Blomqvist,
 \textit{$^{a}$}
 Roberto Cerbino,
 \textit{$^{c}$}
 Caroline Beck Adiels, \textit{$^{a}$}
 Giovanni Volpe, \textit{$^{a}$}

 \\

 & \noindent\normalsize \textit{Myxococcus xanthus} is a unicellular organism whose cells possess the ability to move and communicate, leading to the emergence of complex collective properties and behaviours. This has made it an ideal model system to study the emergence of collective behaviours in interdisciplinary research efforts lying at the intersection of biology and physics, especially in the growing field of active matter research. Often, challenges arise when setting up reliable and reproducible culturing protocols. This tutorial provides a clear and comprehensive guide on the culture, growth, development, and experimental sample preparation of \textit{M. xanthus}. Additionally, it includes some representative examples of experiments that can be conducted using these samples, namely motility assays, fruiting body formation, predation, and elasticotaxis.    \\

\end{tabular}

 \end{@twocolumnfalse} \vspace{0.6cm}]


\renewcommand*\rmdefault{bch}\normalfont\upshape
\rmfamily
\section*{}
\vspace{-1cm}


\footnotetext{\textit{$^{a}$~Department of Physics, University of Gothenburg, SE-41296, Gothenburg, Sweden E-mail: jesus.manuel.antunez.dominguez@physics.gu.se; giovanni.volpe@physics.gu.se}}

\footnotetext{\textit{$^{b}$~
Laboratorio Nacional de Ciencias de la
Sostenibilidad (LANCIS), Instituto de Ecología,
Universidad Nacional Autónoma de México, C.P. 04510, Ciudad de México, México}}

\footnotetext{\textit{$^{c}$~
University of Vienna, Faculty of Physics, Boltzmanngasse 5, 1090 Vienna, Austria}}

\footnotetext{\textit{$^{d}$~Vienna Doctoral School in Physics (VDSP), University of Vienna }}

\footnotetext{\textit{$^{e}$~
Instituto de Física, Universidad Nacional Autónoma de México, C.P. 04510, Ciudad de México, México}}

\footnotetext{\dag~Electronic Supplementary Information (ESI) available: \textbf{}}


\begin{figure*}
\centering
\includegraphics[trim=0 0 0 0, clip, scale=0.35, angle=0, origin=c]{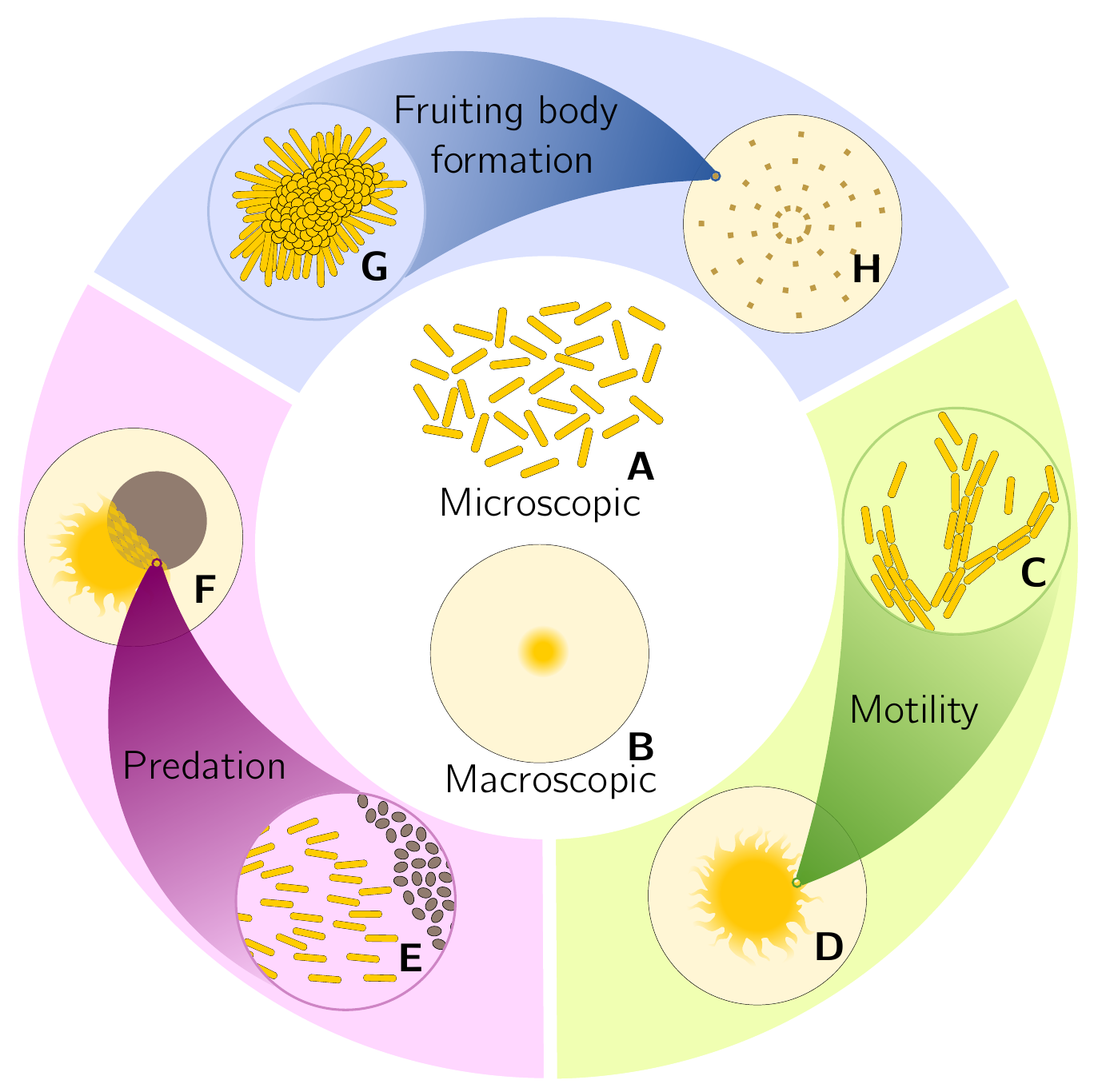}
\caption{{\bf \textit{Myxococcus xanthus} colonies develop different strategies to adapt to their environment, leading to the formation of macroscopic patterns from microscopic entities.} 
{\bf A.} Single \textit{M. xanthus} bacteria (represented by dark yellow rods) spread randomly and moving independenly from each other.
{\bf B.} A freshly inoculated agar plate (light yellow circle) shows no macroscopic pattern because the bacteria are not organised.
{\bf C.} Over time, the bacteria join together into swarms to more effectively explore their environment.
{\bf D.} In the agar plate, the colony develops flares at the edges as a consequence of the swarming.
{\bf E.} When another colony of microorganisms (brown circles)  is found, \textit{M. xanthus} organises itself into ripples to feed on them.
{\bf F.} Rippling can be be seen as macroscopic waves of bacteria in the agar plate. 
{\bf G.} If starved, \textit{M. xanthus} forms mound-shaped fruiting bodies to produce myxospores capable of surviving for long times in very harsh environments.
{\bf G.} The fruiting bodies appear over the agar plate in regular (typically, circular) patterns.}
\label{fig:1}
\end{figure*}

\section{Introduction}

\textit{Myxoccocus xanthus} is a bacterium extensively used in research as a model organism. In Nature, it is found in soils, close to decaying organic matter where it feeds on other microorganisms\cite{Reichenbach1999}. 
It produces antibiotics that allow it to compete with other soil bacteria, and that make it an interesting organism in antibiotic (resistance) studies\cite{Perez2020,Clements-Decker2022}. 
Furthermore, it is able to form myxospores that can endure harsh environmental conditions such as drought, starvation or exposure to chemicals---which makes it interesting for environment sterilisation and bacterial ecology studies\cite{Muller2012,Nehring2015}. 
Interestingly, both these phenomena are closely related to collective behaviours, where \textit{M. xanthus} features a fascinating transition between unicellular and multicellular organisation\cite{Berleman2007,Claessen2014,Mercier2016,AriasDelAngel2020,Ramos2021}. 
    
There are biological, chemical and physical mechanisms behind its organisation\cite{Janssen1977,Dworkin1983,Starruß2012,Rivera-Yoshida2019,Ramos2021}, which features a complex array of strategies that allow it to adapt and thrive in different environmental conditions, as schematically shown in Figure~\ref{fig:1}. 
These strategies are similar to those active matter systems feature in complex and crowded environments\cite{bechinger2016active}. 
In Nature, active matter systems can be found at all scales, from traffic jams to human crowds, schools of fishes, flocks of birds, swarms of insects, plankton communities, and motor proteins in the cytosol. 
There are also artificial examples of active matter like robots, Janus particles, and micromotors. 
Bacterial communities, such as those formed by \textit{M. xanthus}, provide a versatile model system to study active matter.
In spite of their apparent simplicity, bacteria display complex collective behaviours. Some (but not many) strains can be cultured and observed in controlled environments. 

Bacterial collective behaviours arise from two independent processes. 
First, most bacteria are able to extract energy from their environment and use it to displace themselves. For example, this motility allows them to escape hostile or unfavourable conditions, greatly contributing to their survival chances. Different mechanisms exist for bacterial movement---and bacterial species can usually employ more than one, opting for the most suitable depending on environmental conditions\cite{Mitchell2006,Wadhwa2022}. Many of these motility strategies involve collective motion, which is often associated to an increased survival probability of at least one individual in the group. 
Second, bacteria can communicate through the secretion and detection of chemical substances, a phenomenon known as quorum sensing\cite{Waters2005,Duddy2021}. While mainly occurring among cells of the same species, communication between different species is also widespread. What is more, bacteria are known to communicate with other complex living beings, such as plants\cite{Shrestha2020}, fungi\cite{Nasslahsen2022}, and animals\cite{Wu2021,Coquant2021}.

By combining their activity and communication, bacteria develop emergent collective properties. Some examples are found in the coordinated movement of swarms for effective environmental exploration\cite{Be’er2020,Partridge2022}, the realization of tasks that benefit the community rather than the individual\cite{Zhang2023,Pan2023}, and the aggregation into colonies for protection\cite{Grobas2021}.

\textit{M. xanthus} features several complex collective behaviours along its life cycle. 
The cells are elongated and flexible (illustrated by the dark yellow rods in Figure~\ref{fig:1}A) and motile only when on a solid surface such as on an agar plate (light yellow circle in Figure~\ref{fig:1}B). Depending on the density of the population, motility can be independent, called \emph{adventurous} (Figure~\ref{fig:1}A), or inside a swarm, known as \emph{social} (Figure~\ref{fig:1}C). Swarming behaviour can be recognised in the colony as branching structures of cells flowing in the same direction (such as the flares in Figure~\ref{fig:1}D). These processes rely on different cellular mechanisms\cite{Mauriello2010}. 
Another example of collective behaviour can be found in its predatory strategies. \textit{M. xanthus} is an epibiotic predatory bacterium\cite{Berleman2008}: when exposed to colonies of other microorganisms, it approaches them and secretes antibiotics and enzymes to feed on them (Figure~\ref{fig:1}E). The whole population associates and organises in ripples to enhance its efficiency (Figure~\ref{fig:1}F). 
When the conditions are no longer favourable for exploration or predation, the community undergoes a change. Some of the cells turn into myxospores inside protective structures known as fruiting bodies (Figure~\ref{fig:1}G) that distribute throughout the agar surface (Figure~\ref{fig:1}H). The formation of myxospores and fruiting bodies may not beneficial for single cells but to the full community\cite{Wireman1977,Cao2015}. This has been proposed as an example of cooperative behaviour, where the focus of preservation shifts from the individual to the community\cite{Thutupalli2015} as well as an example of a transition from unicellular to multicellular life\cite{Ramos2021}. The myxospores in a fruiting body will germinate into a swarm if the conditions become favourable again, restarting the life cycle. 
    
The variety of collective behaviours along its life cycle has spurred the interest in \textit{M. xanthus}. Nonetheless, laboratory culture of this species requires several steps, including solid cultures in agar plates and liquid cultures in a nutritious medium. Additionally, their survival and development demand specific conditions of sterility, humidity, temperature\cite{Janssen1977}, and light\cite{Martínez-Laborda1990,Pérez‐Castaño2022}. Successfully preparing and properly handling bacterial samples require skills that might be missing in an interdisciplinary laboratory with a focus towards chemistry or physics, which constitutes a potential barrier to the study of these systems by researchers without a strong bacterial biology background.

In this tutorial, we offer a straightforward, step-by-step guide for the culture and growth of \textit{M. xanthus}, tailored for interdisciplinary researchers without a strong background in bacterial biology. Additionally, we provide examples of possible experiments to be performed with \textit{M. xanthus}. Beyond covering the basics, we also provide tips and tricks that often do not find place in published articles, maximising reproducibility.
    
\section{Preparation for microbiology experiments}

In this section, we will provide an overview of the necessary laboratory premises, equipment, materials, and media recipes required to ensure successful and reproducible experimental outcomes.

\subsection{Requirements for the laboratory premises}

\textit{M. xanthus} is a soil dwelling bacterium completely harmless to humans. While it cannot cause human infection, the presence of its spores can contaminate tools and surfaces. Consequently, it is critical to properly sterilise or dispose of contaminated material. Using a flame and wiping the work surfaces and equipment with ethanol at $70\%$ concentration is an effective way of sterilisation. After dipping a glass or metallic tool in the ethanol, the excess alcohol is burned for two minutes in the flame. Metallic tools should turn red hot during the process. Otherwise, tools can be disinfected by soaking them for at least 15 minutes in $70\%$ ethanol. 
 
The \textit{M. xanthus} culture should be handled with gloves in a clean environment to avoid contamination from spores or airborne bacterial species. The working area and surrounding surfaces must be sprayed with $70\%$ ethanol and left to evaporate for at least 30 seconds.  A laminar flow hood or a burner are recommended to maintain a clean working area. The latter will be used in this tutorial to minimise equipment requirements. The list of all the equipment needed for this tutorial can be found in Table~\ref{tbl:equipment}. 
 
To ensure reproducibility, the protocol of this tutorial was also replicated in a second laboratory (the primary laboratory is at the Univeristy of Gothenburg, while the second laboratory is at the University of Vienna), where different equipment was employed (see Supplementary Table~\ref{tbl:equipment}). In particular, in this second laboratory neither a laminar flow hood nor a burner were used but a PCR UV Cabinet instead.

\begin{table}[h]
\small
  \caption{{\bf Equipment.}
  List of equipment used in the described protocol.}
  \label{tbl:equipment}
  \begin{tabular*}{0.48\textwidth}{@{\extracolsep{\fill}}ll}
    \hline
Equipment & Specifications  \\ \hline
Autoclave        &  Benchtop Astell autoclave (AMB220) \\
Spectrophotometer    &    WPA Spectrawave S800 Diode Array        \\ 
Shaker incubator    &    KA KS 3000 i control        \\ 
 Vortexer      &    Heidolph Reax top      \\
 Centrifuge   &   Eppendorf MiniSpin\textsuperscript{\textregistered} plus           \\ 
-80\degree{C} freezer    &  Heto/Icebird Mini Freeze 80\\
Water Purification system   &  Milli-Q\textsuperscript{\textregistered} IQ700 \\ 
 Scale        &    Sartorius CP 22025     \\
 Pipettes 1~mL and  $100\,{\rm \upmu L}$  &   Biohit m1000/m100 \\ 
 pH-meter      &    Fisherbrand\textsuperscript{\texttrademark} accumet\textsuperscript{\texttrademark} \\ & AB150 pH Benchtop Meters      \\
 Inoculating loop   &   Microstreaker VWR (391-0471) \\ 
 Burner      &    Usbeck 1422  \\
 $70 \%$ Ethanol(spraying bottle)   &   Polyethylene, M   \\
1~L Screw-top glass bottle   &  BRAND\textsuperscript{\textregistered} Sigma-Aldrich(BR122562)  \\
100~mL Erlenmeyer glass flask   &  Duran\textsuperscript{\textregistered} Sigma-Aldrich(Z232890)  \\

    \hline
\end{tabular*}
\end{table}

\subsection{Materials and medium recipes}

\begin{table}[h]
\small
  \caption{{\bf Consumables.}
  List of consumables used in the described protocol.}
  \label{tbl:comsumables}
  \begin{tabular*}{0.48\textwidth}{@{\extracolsep{\fill}}ll}
    \hline
Consumable & Specifications  \\ \hline
Petri dishes        & Disposable, 55 mm diameter,  \\
& Polystyrene (PS), no vents\\
 Autoclave tape   &    Comply\textsuperscript{\texttrademark} VWR (817-0118) \\
 Parafilm \textsuperscript{\textregistered} M        &    Amcor      \\
 Sterile syringe filters   &   $0.22\,{\rm \upmu m}$ pore size, Luer connection, \\
  & Polytetrafluoroethylene membrane, \\
  & Sartorius, VWR (611-0696) \\
Sterile 3-part syringes      &  5 or 10~mL, Luer connection,   \\
 & Terumo \textsuperscript{\textregistered} (MDSS05SE, MDSS10SE) \\
Sterile pipettes tips   & Sarstedt (70.33030.100)\\ 
1~mL and  $100\,{\rm \upmu  L}$ & \\
 Pure Cotton        &    Webril \textsuperscript{\textregistered} handi-pads      \\
 non-woven fabric gauze   &   Leukoplast \\
 Aluminium foil       &    Universal      \\
 Eppendorf tubes   &   Microtube, 1.5~mL Sarstedt (72.690.001) \\ 
 Cryogenic vials      &    Disposable, 2~mL, Nalgene,       \\
  & self standing, lip seal, VWR (479-1220) \\
 Liquefied Petroleum    &   Biltema (17-977) \\ 
 Gas container 900~mL  & \\
 Wipers       &    Polyester/cellulose, VWR (115-0031)      \\
 Spectrophotometer cuvettes    &  Disposable, 3~mL, 10~mm path, PS,  \\ 
 & Sarstedt (D-51588) \\
 Gloves       &    Disposable, nitrile, VWR      \\
 \hline
\end{tabular*}
\end{table}

\begin{table}[h]
\small
  \caption{{\bf Reagents.}
  List of reagents used in the described protocol.}
  \label{tbl:reagents}
  \begin{tabular*}{0.48\textwidth}{@{\extracolsep{\fill}}ll}
    \hline
Reagent & Specifications\\ \hline
Tris-HCl    &    Sigma-Aldrich (10812846001)        \\
Agar  & Sigma-Aldrich (05040) \\
 $\text{MgSO}_4$.   &   Sigma-Aldrich (M2643)           \\
 $\text{KH}_2\text{PO}_4$    &  Sigma-Aldrich (P5655) \\
$\text{HCl}$    &  Sigma-Aldrich (H9892) \\
$\text{NaOH}$    &  Sigma-Aldrich (S2770) \\
Bacto\textsuperscript{\texttrademark} Casitone        &   Gibco\textsuperscript{\texttrademark},  Life Technologies Corporation \\
 & ThermoFisher (BD 225930)  \\

 Glycerol     &    Sigma-Aldrich (G5516)      \\

    \hline
\end{tabular*}
\end{table}

\begin{table}[h]
\small
  \caption{{\bf Stock solution recipes.}
  Recipes of stock solutions of solid reagents to use in media preparation, namely CTT broth (liquid culture medium that promotes growth) and TPM buffer (liquid solution to induce starvation).}
  \label{tbl:recipes}
  \begin{tabular*}{0.48\textwidth}{@{\extracolsep{\fill}}lll}
    \hline
Stock reagents & Target  & Quantity for 1 L  \\ \hline
TrisHCl        & 0.1 M                & 15.76 g               \\
$\text{KH}_2\text{PO}_4$         & 0.1 M                & 13.61 g              \\
$\text{MgSO}_4$          & 0.1 M                & 24.65 g             \\ \hline
CTT medium (CasiTone Tris\cite{kaiser1977}) & Target  & Quantity for 1 L \\ \hline
Tris-HCl          & 10 mM                & 100 mL (0.1 M)    \\
$\text{KH}_2\text{PO}_4$            & 1 mM                 & 10 mL (0.1 M)     \\
$\text{MgSO}_4$             & 8 mM                 & 80 mL (0.1 M)     \\
Casitone    & 1$\%$                 & 10 g              \\
Milli-Q water      & -                    & 810 mL            \\ \hline
TPM buffer (Tris Phosphate Magnesium) & Target  & Quantity for 1 L \\ \hline
Tris-HCl          & 10 mM                & 100 mL (0.1 M)   \\
$\text{KH}_2\text{PO}_4$            & 1 mM                 & 10 mL (0.1 M)    \\
$\text{MgSO}_4$             & 8 mM                 & 80 mL (0.1 M)    \\
Milli-Q water      & -                    & 810 mL          \\ \hline
\end{tabular*}
\end{table}

All consumable items required are listed in Table~\ref{tbl:comsumables}, while all reagents and media recipes are in Tables~\ref{tbl:reagents} and \ref{tbl:recipes}, respectively. Similar tables for the experiments performed in the second laboratory are provided in Supplementary Information (Table 1). 

Tris Phosphate Magnesium (TPM) buffer is a solution used to keep the pH stable during bacterial development. Since it lacks any nutritional value but provides a stable environment, it is used  to induce starvation conditions. 
TPM is used in liquid form only during sample preparation, and as agar plate during most experiments to induce starvation, encouraging motility, predation and fruiting body formation. 

For culture and growth of \textit{M. xanthus}, we use CasiTone Tris (CTT) medium instead, which has the same composition as TPM buffer with added peptone casitone for nutrition. A peptone is the result of the partial breakdown of a protein into its constituent amino acid chains. In the case of casitone, the protein casein is broken by animal pancreatic enzymes into chains of different sizes. The result is a supplement with variable composition, as opposed to a chemically defined medium, but with all the necessary components for bacterial growth. CTT is required as both liquid medium and agar plates to culture \textit{M. xanthus}, but it can also be used in motility experiments on agar plates. 

When properly sealed and maintained sterile, the autoclaved media can be stored at room temperature, protected from direct sunlight and large temperature fluctuations.  While TPM buffer can be viable and stored for more than a year, CTT medium should be used within 3 to 6 months. The medium should be clear yellow, hence turbidity  or changes in the colour of the media are signs of degradation and contamination, leading to their disposal. 

To prepare solid media, add agar in the proportion of 1.5$\%$ expressed as mass of solute over volume of solution, that is, 15~g for 1~L.  Before use, the pH has to be adjusted to 7.6 and the media has to be sterilised in the autoclave at 121\degree{C} for 15 minutes. When properly sealed, sterile agar plates,  can be stored in the same conditions as the liquid media for up to 3 months for culture continuation. However, for experiments, fresh agar plates are advised to ensure an accurate water content. A decrease in the level of the agar and the appearance of unknown colonies indicate inefficient sealing of the plates, rendering them inadequate for use.

\subsection{Bacterial strain}

The widely used laboratory wild-type strain DK1622 of \textit{M. xanthus} was specifically employed for the procedures outlined in this tutorial. This strain serves as a robust starting point that may be applicable to other strains, including those isolated from natural environments\cite{Vaksman2015}. However, other strains and related species may exhibit variations in development times, temperature preferences, and growth media. 

\section{Culture protocols}

\begin{figure*}
\centering
\includegraphics[trim=0 0 0 0 ,clip, scale=0.9
, angle=0,origin=c]{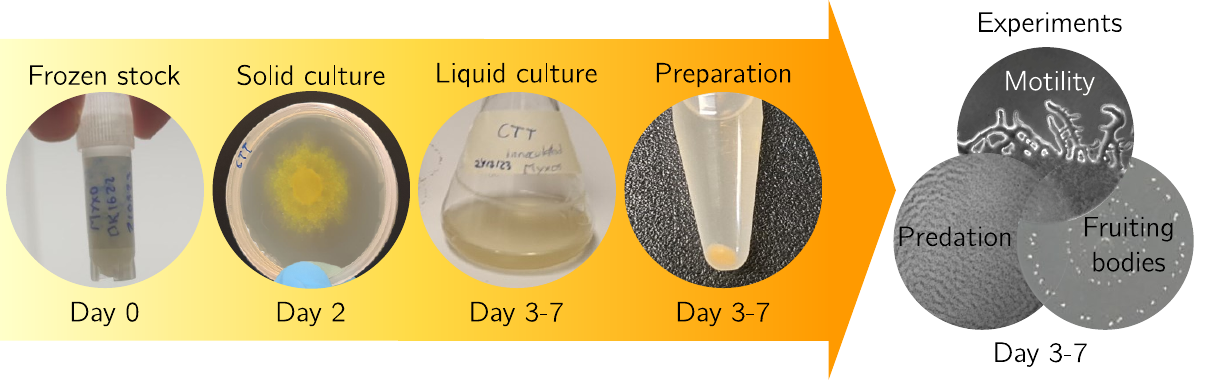}
\caption{{\bf Steps for the culture of \textit{M. xanthus}.} 
From a long term storage glycerol stock, an initial solid culture is required to develop a viable colony. This colony is then transferred to liquid medium for better assessment of cell concentration through Optical Density ($\text{OD}_{600}$) for experiment repeatability during each day of the week. After centrifuging and resuspending, cells can be used in experiments. Here, we focus on three main phenomena: motility of the community and cells, predation of other microbial colonies, and formation of fruiting bodies in adverse environmental conditions. }
\label{fig:2}
\end{figure*}

The complete growth process of \textit{M. xanthus} culture is portrayed in the timeline in Figure~\ref{fig:2}. We propose the weekly schedule described in Table~\ref{Scheduletable}, in which a new colony from a frozen stock is revived at the end of every week to be used in experiments during the following week. Agar cultures are used at the beginning for the recovery of a viable population from the frozen stock. For experiments, liquid culture is preferred to assess population density according to $\text{OD}_{600}$ (optical density at 600~nm illumination wavelength) measurements with reference to a blank sample of sterile CTT medium. Every day of the week, an overnight concentrated liquid culture is used for experiments and as an inoculate for the culture of the following day, resulting in five independent experiments per week. After five days of experiments, the culture is discarded, with a new solid culture inoculated from a frozen stock.

\begin{table*}
\small
  \caption{{\bf Proposed schedule for \textit{M. xanthus} culture.} 
  A new plate is inoculated on the first day to be used during five days of experiments. The ``Requirements'' column indicates what is needed in advance. The ``Activity'' column refers to the relevant sections of this tutorial. The ``Results'' column indicates the outcome of each activity, organised to match the requirements of the next day.}
 \label{Scheduletable}
  \begin{tabular*}{\textwidth}{@{\extracolsep{\fill}}llll}

    \hline
        \textbf{Day } & \textbf{Requirements } & \textbf{Activity } & \textbf{Result } \\ \hline 
        \textbf{In advance } & Material \& Reagents & Preparation of culture media    & CTT plates    \\ 
        & stockage &  &   CTT flasks \\ 
        &  &  &   TPM plates \\  
        \textbf{Day 1}   & CTT plate   & From frozen stock to agar plate  &  Inoculated CTT plate \\
        &  Frozen stock &  &  \\  \\
        \textbf{Day 3} &  Grown colony plate   & From agar plate to liquid medium  & Inoculated CTT flask    \\
        &  CTT flask &  &   \\ \\
        \textbf{Day 4} & CTT flask  & Liquid culture continuation & Inoculated CTT flask    \\
        & TPM plate & Experiments   &  Experimental data \\
        & Overnight culture flask & *If few frozen stocks are left:    & Frozen stocks   \\
        &  & Liquid medium to frozen stocks  &    \\  \\
        \textbf{Day 5} & CTT flask  & Liquid culture continuation & Inoculated CTT flask    \\
        & TPM plate & Experiments   &  Experimental data \\
        & Overnight culture flask & &   \\ \\
        \textbf{Day 6} & CTT flask  & Liquid culture continuation & Inoculated CTT flask    \\
        & TPM plate & Experiments   &  Experimental data \\
        & Overnight culture flask & &   \\  \\
        \textbf{Day 7} & CTT flask  & Liquid culture continuation& Inoculated CTT flask    \\
        & TPM plate & Experiments   &  Experimental data \\
        & Overnight culture flask & &   \\  \\
        \textbf{Day 8} & TPM plate & Experiments & Experimental data    \\
        & Overnight culture flask & &   \\
        & CTT plate   & From frozen stock to agar plate  &  Inoculated CTT plate \\ 
        & Frozen stock &  &    \\ \hline
        
    \end{tabular*}
    
\end{table*}

\subsection{Preparation of liquid media}

Liquid CTT is used during the culturing steps of M.xanthus, so it is important to maintain a fresh supply available at all times and renovate it every 3 months. Liquid TPM is needed only during sample preparation, but it can be stored for more than a year. Preparing a litre of medium is recommended for continued use, but it can be adjusted according to experimental needs. 

\begin{enumerate}

\item Mix the ingredients of the corresponding medium recipe in Table~\ref{tbl:recipes}. Scale accordingly depending on the volume to be prepared.
\begin{enumerate}
\item If using casitone, mix the powder and solution by shaking until a homogeneous solution is reached.
\item[] NOTE: It is also possible to mix using a magnetic stirrer.
\end{enumerate}

\item Adjust the pH of the solution to 7.6 adding 1 M of HCl and NaOH with the help of a pH-meter.

\item The medium can then be directly autoclaved to be stored as liquid, added agar to make plates as detailed in Section~\ref{sec:agar} or, in the case of CTT, be poured directly into flasks for liquid cultures as described in Section~\ref{sec:liquid}.

\end{enumerate}

\subsubsection{Autoclaving}\label{autoc}

Autoclaving is required to ensure sterility of media. It can also be used with other compatible objects like autoclavable cryogenic vials, flasks, or glass bottles

\begin{enumerate}

\item Place the medium in a glass container.

\item  Screw the cap of a screw top bottle container until it is stable and almost fully tight, but still allowing for air exchange. Put a piece of autoclave tape connecting the flask and the cap. The change in colour of the tape after the autoclave cycle will ensure sterilisation was successful.

\item  Autoclave the media for 15 minutes at 121\degree{C} (consider that the process takes about 2 hours because the autoclave has to reach the temperature, and once the desired time has passed, it has to cool down and depressurise).

\item When cooled down and safe to touch open the chamber and tightly close all the recipient caps before storing or transferring to a disinfected surface.

\item [] CAUTION: Do not try to autoclave plastic
petri dishes.
\end{enumerate}

\subsubsection{Preparation of the agar plates}
\label{sec:agar}
CTT agar plates are essential for \textit{M. xanthus} cultures, with a minimum of one used every week during experiments. TPM agar plates are the substrate of most experiments, so it is convenient to store them in large numbers.
As an approximation, 7~mL of medium should be used per 55~mm diameter plate (even though anything between 5 and 10~mL is adequate). Planning for their use in 3 months, approximately 105~mL of CTT will produce 15 agar plates. As for TPM, 455~mL will yield 65 plates to be used in a same 3 months period of experiments. These numbers can be adjusted according to experimental needs.
\begin{enumerate}

\item Pour the desired volume of medium (CTT or TPM) in a glass bottle with at least double the capacity.  

\item  Add 1.5$\%$ (w/v) of agar powder to the medium in the glass container.

\item  Autoclave the mixture according to the steps in the section~\ref{autoc}.

\item  Transfer the recipient while still hot, but safe to the touch, to a clean environment, such as a laminar flow hood or in the vicinity of a lighted burner, as shown in Figure~\ref{fig:3}A. In the absence of both, as in the second laboratory, we observed occasional sample contamination that was avoided only by restricting laboratory access to the smallest number of personnel and closing the cabinet door in between steps.

\item  Pour the agar on the Petri dishes until it covers 2/3 of the surface. Tilt the dishes to have the agar evenly spread.

\item  Leave the plates open and let them dry. This takes approximately 30 minutes if you are working in a laminar flow bench or more if using a burner depending on the laboratory conditions. 

\item[] NOTE: In the second laboratory, where a closed UV cabinet was used to avoid external airflow without laminar flow or a burner, it took approximately one hour for the droplet to dry.

\item  After drying, close the plates by placing a parafilm strip all around the perimeter covering the space between lid and plate. Store them upside down, to avoid condensation on top of the agar, at room temperature in a clean and dry environment.

\end{enumerate}

\subsubsection{Preparation of Erlenmeyer flasks with liquid media}\label{sec:liquid}

Flasks with TPM are required for the culture of \textit{M. xanthus} for every day of planned experiments. A batch with 5 of them is recommended per week consuming 125~mL of CTT. Flasks should not be stored for more than 2 weeks, since they are not completely sealed and will lose water content over time. 

\begin{enumerate}

\item Prepare caps for the Erlenmayer flasks. 
All the required materials that are shown in Figure~\ref{fig:3}B. Sterility of the materials is not required since they will be autoclaved with the flasks. These caps allow the flasks gas exchange with the environment without being at risk of contamination. 

\begin{enumerate}
    \item Separate a ball of cotton approximately 3~cm of diameter. 
    \item  Cut two gauze squares of approximately 10~cm.  
    \item Put the ball of cotton in the middle of the two layers of gauze and close it as shown in Figure~\ref{fig:3}C.
    \item Secure with masking tape as in Figure~\ref{fig:3}D.
    \item Check that the cap fits the flask's opening tightly.
\end{enumerate}

\item  Pour 25~mL of CTT medium in each of the Erlenmeyer flasks.

\item Place the cap to close the Erlenmeyer flask.

\item Cover the cap with aluminium foil loosely to allow gas exchange. 

\item Sterilise by autoclaving at 121\degree{C} for 15 minutes.

\item Remove the aluminium foil and put parafilm in the gap between the cap and the flask, as the cap will ensure the liquid to stay sterile.

\item Store in a dry and clean space at room temperature until you need to use them.

\end{enumerate}

\subsection{From frozen stock to agar plate}

Frozen stocks are used for long term storage of bacterial strains, lasting for more than a year while kept at $-80$\degree{C}. To recover a healthy viable colony of \textit{M. xanthus}, they are first inoculated in an CTT agar plate at optimal conditions.

\begin{enumerate}

\item  Prepare the work surface by spraying $70\%$ ethanol and spreading it evenly with a wipe.

\item Turn on the burner. It will ensure the sterility of the environment around the flame.

\item Take a sterile CTT medium agar plate, remove the parafilm protection in the vicinity of the flame and place the agar plate on the work surface close to the flame. If there is moisture on the lid, let it dry  by opening it until it disappears. Alternatively, the closed agar plates can be left in the incubator upside down 15 minutes before use to evaporate any condensation. If there are still big droplets on the lid, clean with a disposable wipe damped with $70\%$ ethanol, making sure it completely dries in a clean zone around the flame.

\item Take a frozen glycerol stock from the -80\degree{C} freezer and let it thaw until a small amount of liquid can be seen at the top.

\item Take $50\,{\rm \upmu L}$ of the thawed liquid from the frozen stock of \textit{M. xanthus} using a pipette.
\begin{enumerate}
\item Try to be as fast as possible to avoid further thawing of the glycerol stock before returning it to the freezer. The stocks can be used multiple times for inoculation, but their viability decreases with every thaw---freeze cycle.
\end{enumerate}

\item Place the liquid as a single droplet in the centre of the agar plate as shown in Figure~\ref{fig:3}E and let it dry.

\item If the burner is on, the agar plate, as well as its lid, can be left uncovered in the clean region close to the burner. Placing them behind the burner ensures they are out of the range of movements to be performed during other manual work. Be careful to not tilt the plate to avoid spreading the droplet.

\item  If the burner is off, the lid should be placed half-tilted on top of the plate to allow the moisture to escape faster. It is advised to do so in a corner or surface with no nearby transit. Similarly, a previously disinfected bigger box (plastic or glass) can be placed to cover the drying agar plates from airborne contamination if transit or movement is expected around the drying environment.

\item Once the droplet is fully dried and is barely visible, close the agar plate with the lid and place a parafilm strip around the seal. 

\item  Place the agar plate upside down in an incubator at 32\degree{C}, protected from direct light exposure. Wrap the agar plate with aluminium foil if the incubator has a window.

\item  The colony will be ready to be suspended for liquid culture after two days of growth. Figure~\ref{fig:3}F shows how the colony looks like after 48 hours of incubation. 

\end{enumerate}

\subsection{From agar plate to liquid medium}

In the agar plate, the bacteria will grow and spread unevenly over the surface. Therefore, solid support is inconvenient for extraction and quantification of cells. Liquid cultures can be monitored for optical density, that relates to cell concentration. However, \textit{M. xanthus} is unable to swim and the culture must be shaken to maintain the bacteria in suspension.

\begin{enumerate}

\item Secure a clean area by  spraying 70$\%$ ethanol and spreading it evenly with a wipe.
    
\item Turn on the burner. This will ensure the sterility of the environment around the flame.

\item Take a sterile CTT flask and remove the protective parafilm.

\item Take a CTT agar plate inoculated with \textit{M. xanthus} and grown for 48~h out of the incubator. Keep it upside down in the working environment.

\item Remove the protective parafilm around the seal of the agar plate once inside the clean area. 

\item Place the agar plate facing up without the lid in the clean zone.

\item Sterilise the inoculating loop by spraying it fully with 70$\%$ ethanol. 

\item Once the ethanol is fully evaporated, spray only the tip of the loop with 70$\%$ ethanol and pass it through the flame until it turns red hot. 

\item Allow the loop  to cool down by gently shaking it in the air close to the burner.

\item Check the temperature of the loop by taking the grown CTT plate and touching a peripheral area of the agar far from the \textit{M. xanthus} colony. It should not melt or make a crackling noise when in contact with the agar.

\item[] NOTE: Skip steps 8 to 10 if using plastic disposable loops.

\item Once cooled, scrape the whole colony as shown in Figure~\ref{fig:3}G.  

\item Remove the cap from the CTT flask, and while keeping it on one hand in the clean area, disperse the colony in the CTT liquid medium using the other hand. See Figure~\ref{fig:3}H.

\item Place the cotton cap back on the opening of the CTT flask. Do not secure it with parafilm to allow air circulation through the cotton cap.

\item Place securely on an incubator with a shaker overnight at 32\degree{C} and at 250~rpm and in the dark. If the incubator has a transparent window, wrap the flask with aluminium paper once inoculated to avoid exposure to light. Figure~\ref{fig:3}I shows the wrapped flask with the horizontal plastic holders that secure it in place while shaking.

\end{enumerate}

\subsection{Liquid culture continuation}

\textit{M. xanthus} in liquid culture will actively grow for limited period of time. Over time, the community will stagnate due to nutrient depletion while accumulating a significant amount of dead cells and waste material. This will affect the viability of the samples and reliability of concentration measurements. To ensure the repeatability of experiments, a new liquid culture is started to be used on experiments the next day.

\begin{enumerate}

\item Secure a clean area by  spraying 70$\%$ ethanol and spreading it evenly with a wipe.

\item Turn on the burner. It will ensure the sterility of the environment around the flame.

\item  Open a fresh CTT liquid medium flask by removing the protective parafilm.

\item Take 3~mL of undiluted overnight culture with a pipette as inoculate. See the contrast between an overnight culture and a sterile one in Figure~\ref{fig:3}J.

\item  Add the inoculate into a fresh CTT liquid medium flask.

\item Place securely on an incubator with a shaker overnight at 32\degree{C} and at 250 rpm and preferably in the dark.

\end{enumerate}

\subsection{Liquid media to frozen glycerol stocks}

After obtaining successful growth of any strain of bacteria, it is convenient to secure a viable population for future experiments. Here, we include instructions for long term storage by making glycerol stocks. The resulting glycerol stocks are ready to use and can last for years. Each one can be used multiple times; nevertheless, each thaw-frezee cycle will decrease the viability of the extracted inoculate.

Glycerol stocks of the same batch come from a homogeneous population and are less susceptible to variations. Therefore, it is recommended to produce many glycerol stocks of the same culture to be stored and used over long periods of time rather than frequent production of stocks that might accumulate variations over time.

Glycerol stocks are the preferred means of shipping if kept at $-80$\degree{C}. Otherwise, it is preferable to send recently inoculated agar plates that must be transferred to liquid medium on the day of arrival.

\begin{enumerate}

\item Incubate and shake, at 32\degree{C} and at 250~rpm, a liquid culture of \textit{M. xanthus} in a CTT medium flask until the $\text{OD}_{600}$ measurement reaches between 1.5 to 2.0. This should correspond to roughly two to three days of culture.

\item Prepare a dilution of 50$\%$ (v/v) glycerol in CTT. Mix $250\,{\rm \upmu L}$ of glycerol and  $250\,{\rm \upmu L}$ of CTT for each sterile cryotube to be filled. Autoclavable cryotubes can be previously sterilised using the autoclave if they are not taken from sterile packaging according to the steps in Section ~\ref{autoc}. 

\item Filter the glycerol solution using a syringe and passing it through a $0.22\,{\rm \upmu m}$ pore size filter into a sterile container in a clean environment. If made in excess, this freezing rich medium can be stored at room temperature if kept sterile and air tight.

\item Fill each of the cryotubes by placing $500\,{\rm \upmu L}$ of the glycerol dilution first.

\item With a sterile pipette, place $500\,{\rm \upmu L}$ of the bacterial suspension inside the cryotubes.

\item Close the cryotubes, label them and place them in a -80\degree{C} freezer.

\end{enumerate}

\subsection{Washing samples for experiments}

Having obtained overnight liquid cultures of \textit{M. xanthus} in CTT, samples must be prepared to be used. The CTT medium leftovers must be washed to not interfere with the starvation conditions of most experiments. Additionally, the population density needs to be assessed through the OD for repeatability.

\begin{figure*}
\centering
\includegraphics[trim=0 0 0 0 ,clip, scale=0.7
, angle=0,origin=c]{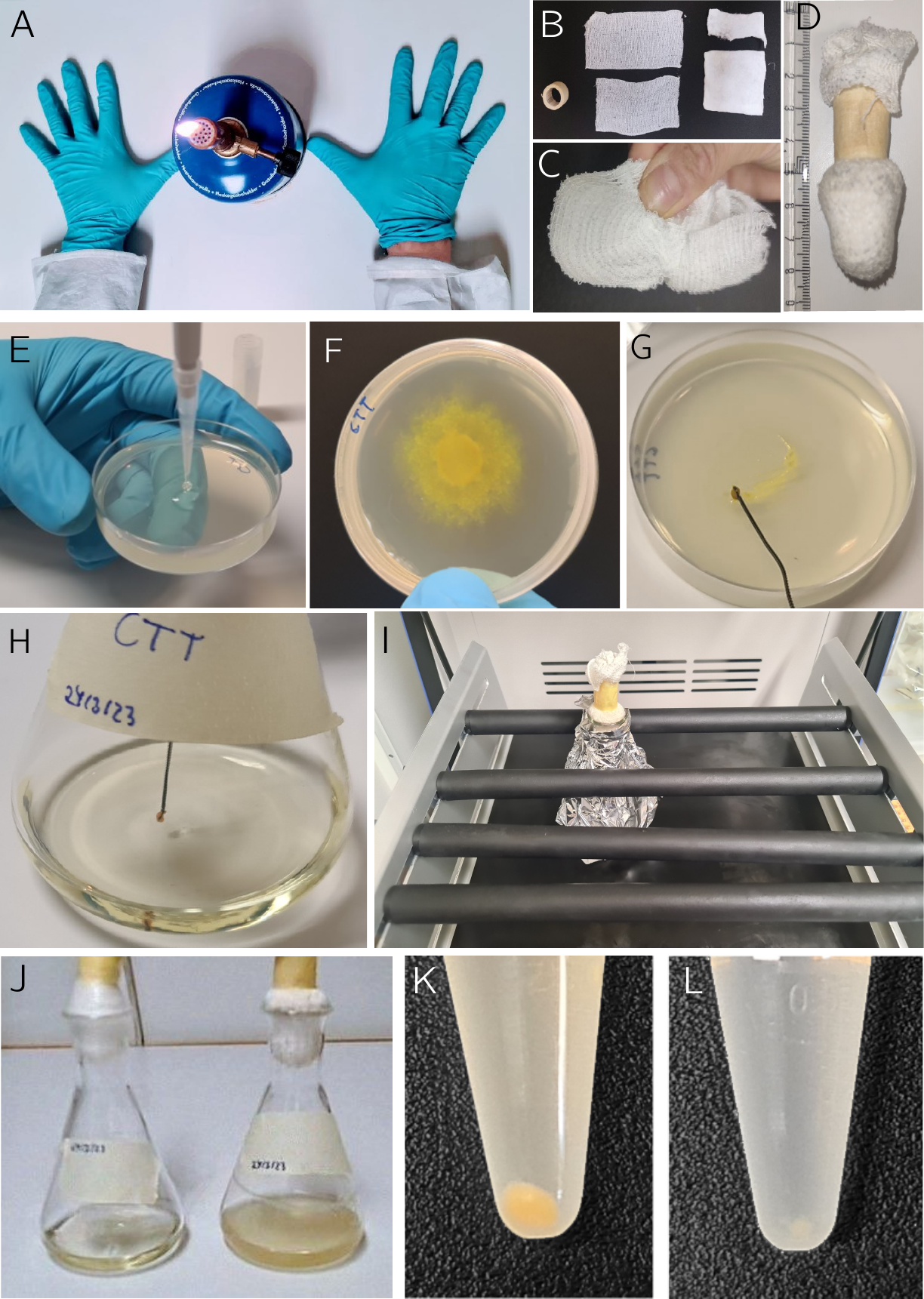}
\caption{{\bf Main steps for the culture of \textit{M. xanthus}.} 
{\bf A.} Approximation of the clean working space provided by a burner. 
{\bf B.} Materials used in the fabrication of cotton caps for liquid cultures, including gauze, cotton and tape. 
{\bf C.} Assembly of the materials of the cotton cap. The cotton is shaped into a ball, placed in the middle of two superposed squares of gauze and wrapped tightly by joining all the edges and corners of the gauze. 
{\bf D.} Finished cotton cap after securing with tape. 
{\bf E.} Inoculation of a CTT agar plate with \textit{M. xanthus} from a frozen stock. 
{\bf F.} CTT agar plate with a colony grown for 48 hours. 
{\bf G.} Scrapping the full colony using an inoculation loop. 
{\bf H.} Inoculation of liquid medium with a colony from a CTT plate. 
{\bf I.} Placement of the liquid culture flask in the shaker incubator, covered with aluminium foil. 
{\bf J.} Comparison of turbidity of an sterile CTT medium flask versus a proliferated CTT flask ready to use in experiments. 
{\bf K.} Pellet of a centrifuged sample of \textit{M. xanthus} for dense population experiments ($\text{OD}_{600} = 0.5$). 
{\bf L.} Pellet of a centrifuged sample of \textit{M. xanthus} for sparse population experiments ($\text{OD}_{600} = 0.05$).}
\label{fig:3}
\end{figure*}

\begin{enumerate}
    
\item Extract 1~mL of the cultivated medium with \textit{M. xanthus} and place it in a sterile spectrophotometer cuvette in a clean environment. 

\item Measure the $\text{OD}_{600}$ of the overnight culture.

\item Adjust the concentration by extracting bacterial suspension and diluting with clean CTT until the desired concentration is reached. Here, two population densities are considered. Dense population samples have an $\text{OD}_{600}$ of 0.5, while the sparse population samples have an $\text{OD}_{600}$ of 0.05. However, this parameter can be adapted and tuned according to the need of each experiment. In Figures~\ref{fig:3}K and \ref{fig:3}L, see the contrast  between the resulting pellets of a concentrated and diluted sample, respectively. In the latter, the pellet might not be visible by naked eye.

\item Put 1~mL of the cultivated media with \textit{M. xanthus} in the desired concentration in an Eppendorf tube.

\item Centrifuge for 5 minutes at 8000~rpm ($4300 g$ in relative centrifugal force).

\item Open the lid of the Eppendorf tube in a clean area and get rid of the supernatant until only left with the pellet in the bottom of the tube. 

\item Add 1~mL of liquid TPM to the Eppendorf tube.

\item  Close the lid of the tube and resuspend the pellet with the vortex or pumping the liquid up and down with a sterile pipette.

\item Centrifuge again for 5 minutes at 8000~rpm ($4300 g$).

\item Open the lid of the Eppendorf tube in a clean area and get rid of the supernatant until only left with the pellet in the bottom of the tube. 

\item Add $100\,{\rm \upmu L}$ (10th of the original volume of the liquid) of liquid TPM. 

\item Resuspend with the vortex or pipette. The resulting suspensions is now ready to be used in experiments.

\end{enumerate}

\section{Experiments}

This section showcases some experiments that can be performed with \textit{M. xanthus} to give an overview of the potential research topics that can be studied with this model organism. Each of the experiments described below is independent, but it is possible to perform multiple experiments in the same agar plate if separated by enough space.

(Note that the detailed specifications for experiment design are beyond the scope of this paper. In particular, the image acquisition procedures will depend on the specifics of each optical setup (for example, the use of an inverted or non-inverted microscope, or the access to temperature and humidity control with a stage incubator). The main challenges to overcome are condensation in the optical window, agar water loss over time, and the lack of agar surface flatness.)

\subsection{Motility}

\begin{figure*}
\centering
\includegraphics[trim=0 0 0 0 ,clip, scale=0.72
, angle=0,origin=c]{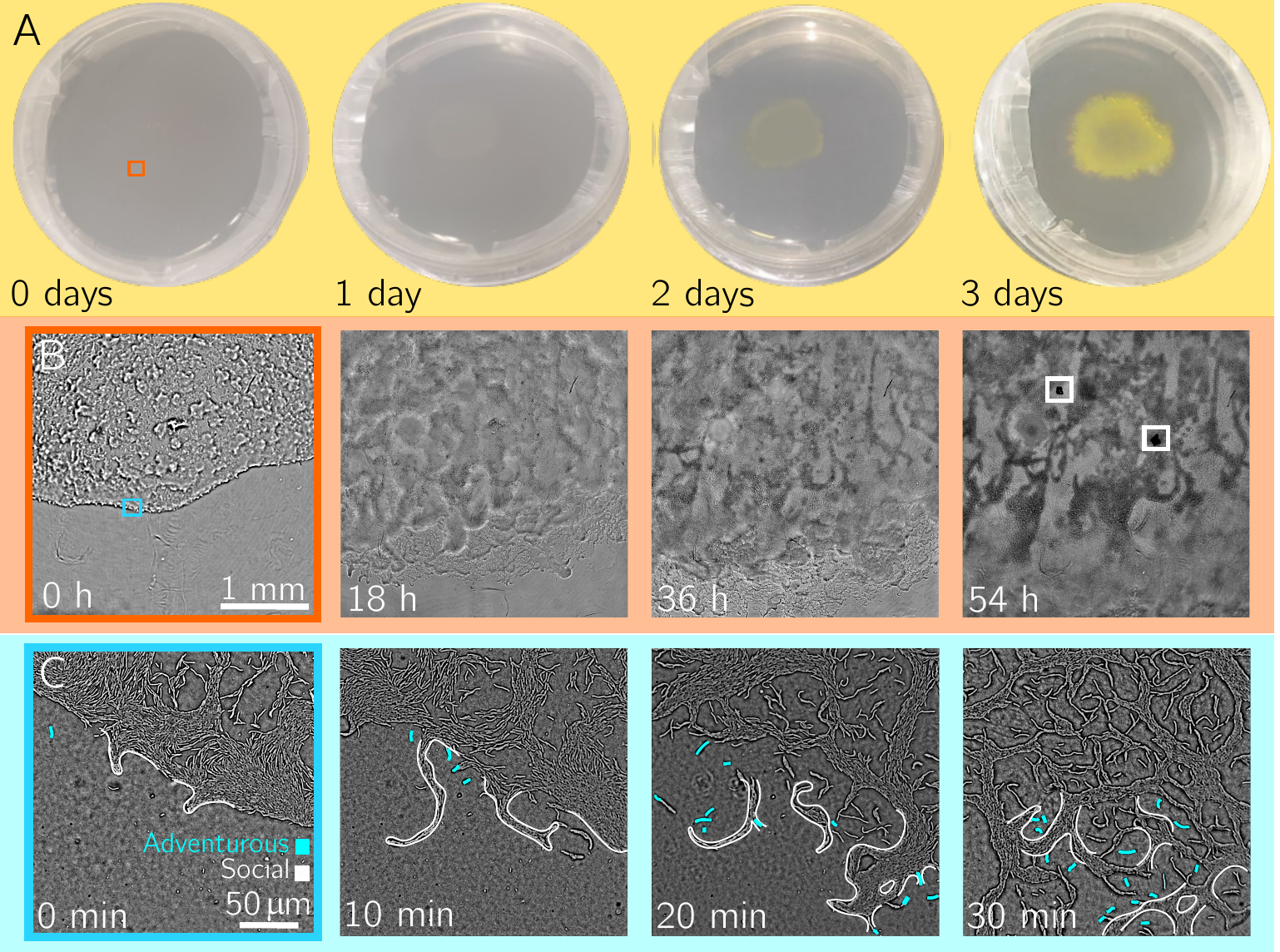}
\caption{{\bf \textit{M. xanthus} motility at different spatial and temportal scales.} 
{\bf A.} Full colony behaviour on a CTT agar plate followed over the course of 3 days at 32\degree{C}, where the production of pigment intensifies and the swarm spreads through the plate. The full Petri dish (55~mm diameter) is shown over a black background to highlight the colonies. 
{\bf B.} Section of the community followed for 54 hours to detect the displacement of swarms. Darker zones are due to pigment accumulation and struvite crystals are highlighted at 54 hours. Images at $4\times$ magnification.
{\bf C.} Microscopic view of a monolayer of cells moving forward in the course of 30 minutes. The two types of motility as highlighted at the edge of the colony. Social motility leads to the formation of flares, while events of adventurous motility can be identified in single scout cells. The images were taken at $63\times$ magnification.
}
\label{fig:4}
\end{figure*}

\textit{M. xanthus} requires a solid surface for displacement. Depending on the spatial scale of observation, different time scale should also be considered. 
Looking at the full colony development on CTT agar plates on the course of days, one can observe pigment production (DKxanthene\cite{Meiser2006,Meiser2008}) as well as displacement of swarms from the original inoculation site, as shown in Figure~\ref{fig:4}A. 
In the presence of light, different pigments will be produced and \textit{M. xanthus} growth will be affected\cite{Martínez-Laborda1990}. In contrast, in TPM agar plates, no pigment is produced and the colony remains almost invisible to the naked eye. 

The collective behavior of a section of the bacterial colony can be monitored for hours to analyze the displacement of swarms, as shown in Figure ~\ref{fig:4}B. After 48 hours of incubation on CTT plates, regions of pigment accumulation and crystal formation become visible (see Supplementary Video 1). These crystals are struvite, formed due to the precipitation of the ammonia waste produced by \textit{M. xanthus}  whith the magnesium ions and phosphate in CTT. M. xanthus is known for its ability to biomineralise and precipitate different minerals depending on the medium composition\cite{Jimenez-Lopez2008,González-Muñoz2010}. This phenomenon has been utilized in practical applications such as bioremediation of nuclear elements and the protection of sculptures\cite{Rodriguez-Navarro2003,Kataki2019}.

At the microscopic scale, the organisation and orientation of each cell can be observed over minutes. Two kinds of motility come into focus. 
When isolated, cells move according to adventurous motility, while in high concentration, social motility will be predominant leading to the formation of swarms\cite{Dinet2023}(see Supplementary Video 2 and 3, respectively). In the latter case, cells will form swarms leading to expanding fronts in the colony known as flares\cite{Dworkin1983,Koch2006,Ritchie2021} seen in Figure~\ref{fig:4}C (and Supplementary Video 4 and 5). Still, some individual cells might sometimes leave the swarm to explore the environment through adventurous motility.  The rigidity of the medium affects both motility mechanisms and so, different concentrations of agar will yield different results\cite{Patra2016,Rivera-Yoshida2019}.

This is the protocol to realize these experiments:
\begin{enumerate}

\item Place a droplet of $13\,{\rm \upmu L}$ of \textit{M. xanthus} of the desired concentration in the  surface of a TPM or CTT agar plate and let it dry.  
\begin{enumerate}
\item For the macroscopic case, take images every hour for a total duration of 48 hours.
\item For the microscopic case, take images every 2-10 minutes for a total duration of 24 hours. 
\end{enumerate}

\item[] NOTE: The suggested times are only an orientation and development might still vary from one experiment to another.
    
\end{enumerate}

\subsection{Fruiting bodies}

\begin{figure*}
\centering
\includegraphics[trim=0 0 0 0 ,clip, scale=1
, angle=0,origin=c]{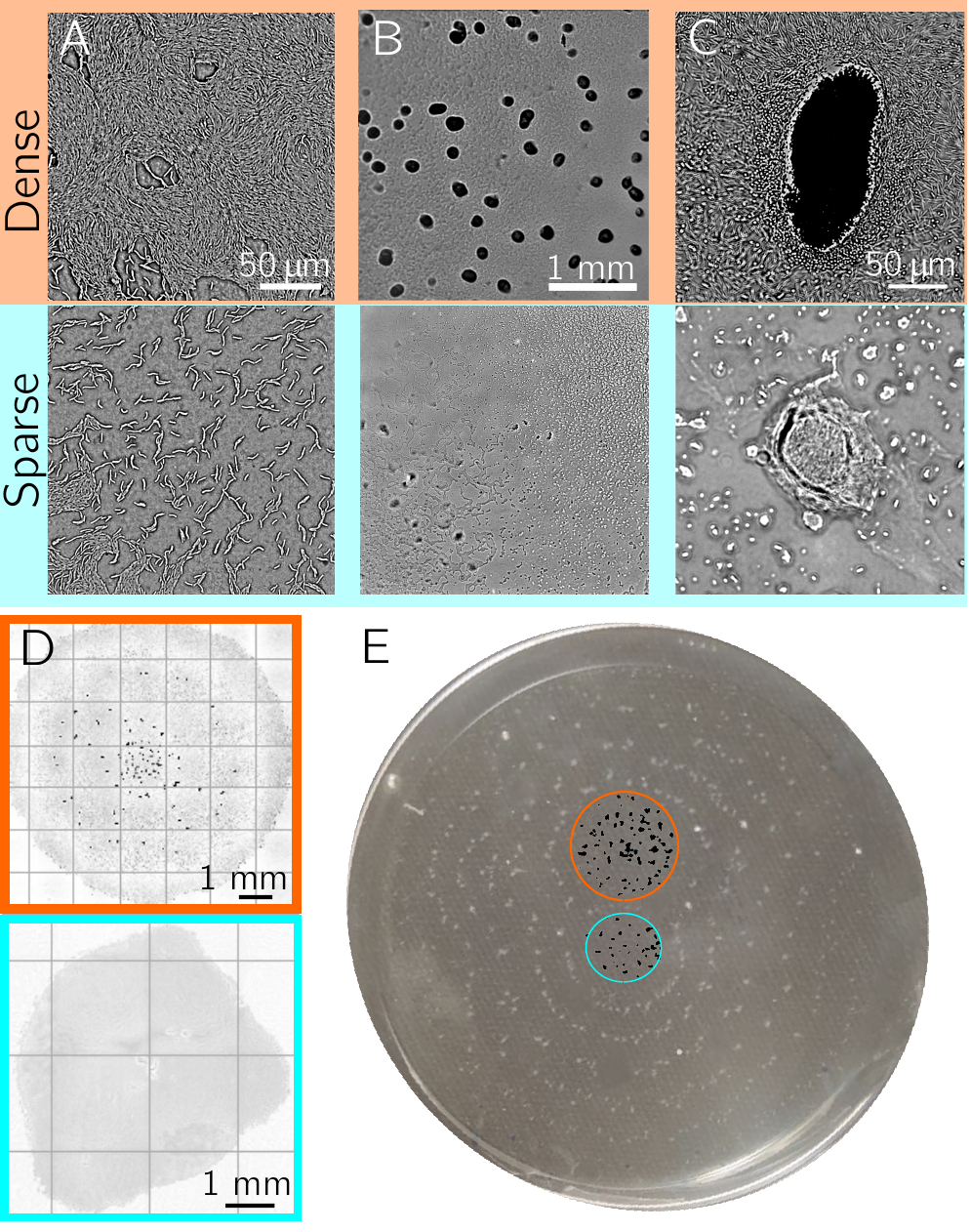}
\caption{{\bf Fruiting body formation}
{\bf A.} Dense and sparse populations on TPM agar plates at the initial time point observed at $63\times$ magnification. 
{\bf B.} Fruiting body formation at $4\times$ magnification after at 48 hours.  
{\bf C.} Close-up of a fruiting body at $63\times$ magnification.
{\bf D.} Comparison of the colony spread and fruiting body formation at 48 hours depending on the initial population density. 
The images result from the merging of adjacent pictures at $4\times$ magnification.
{\bf E.} Concentric pattern of fruiting bodies after two weeks incubation. The fruiting bodies of dense and sparse populations before both made contact are highlighted in black. 
The initial inoculation site of each population is at the center of the two circles. The full Petri dish of 55~mm of diameter is shown against a dark background to emphasise the fruiting bodies.}
\label{fig:5}
\end{figure*}

Fruiting bodies are defensive structures where the cells of \textit{M. xanthus} congregate and some of them differentiate into myxospores in a phenomenon known as sporulation\cite{Meiser2006,Bretl2016,Muñoz2016}. Spores are impervious to extreme conditions and ensure community survival if the environment turns hostile. Fruiting body formation occurs during the natural development of \textit{M. xanthus} but their formation can also be triggered by different means\cite{Kimura2010,Müller2012,Yoshio2022}: a decrease in humidity, an increase of certain chemical species, such as glycerol, or as in the example of this tutorial, starvation. 

In this tutorial, we will show that density population has a significant impact on the formation of fruiting bodies, which is an example of collective behaviour.
We compare two different population densities, designated as ``dense'' and ``sparse'', with an $\text{OD}_{600}$ of 0.5 and 0.05 respectively, which are shown in Figure ~\ref{fig:5}A. 
In the dense population, cells are evenly distributed in layers\cite{Copenhagen2021} (see Supplementary Video 6) with few empty spaces, while in the sparse case, most cells are isolated with most of the agar remaining empty.

After 48~h, fruiting bodies are considerably more abundant and developed in the dense population (see Supplementary Videos~7 and 8) as opposed to the sparse one (see Supplementary Video~9), as shown in Figure~\ref{fig:5}B. A microscopic look on individual fruiting bodies allows to identify myxospores, recognised by their round shape (see Supplementary Video 10). They can be found mainly in the centre of the fruiting body stacked on each other, even though there are also myxospores scattered around the nearby area. Vegetative cells, in the shape of elongated rods, are distributed mostly in a layer and around the fruiting body, forming a structure called haystack\cite{Jelsbak1999,Sozinova2006,Muñoz2016} . Not all cells will undergo differentiation in the fruiting body, as some remain to prey on possible targets or even sacrifice to sustain the sporulation process\cite{Wireman1977,Oconnor1988}. 
The fruiting bodies in a sparse population lag in growth and development when compared to those in a dense population, as shown in Figure~\ref{fig:5}C. 

In the sparse colony, adventurous motility will predominate at the beginning, leading to a slower spread of the colony in hard agar (1.5$\%$ w/v), like in Figure~\ref{fig:5}D (while the opposite will be true for soft agar\cite{Patra2016}). 
If incubated for a couple of weeks, \textit{M. xanthus}  fruiting bodies will appear periodically in concentric lines around the original culture\cite{Berleman2007,Rivera-Yoshida2019} as shown in the plate in Figure~\ref{fig:5}E. Such pattern is not present in the inoculation site, that is, the confines of the original inoculated droplet (where the fruiting bodies positions are random), but instead outside, due to the population propagation. 

This is the protocol to realize these experiments:
\begin{enumerate}

    \item Place a droplet of $13\,{\rm \upmu L}$ of \textit{M. xanthus} on the  surface of a TPM agar plate and let it dry. 
    \begin{enumerate}
    \item For the macroscopic case, take images every 1 hour for a total duration of 96 hours and compare the results between different concentrations (OD) of  \textit{M. xanthus}.
    \item For the microscopic case, take images every 10-20 minutes for a total duration of 96 hours.
    \end{enumerate}
    
    \item Keep in mind that fruiting body formation can take up to 48 hours to begin after inoculation. 
    
    \item[] NOTE: The suggested times are only an orientation and development might still vary from one experiment to another.
    
    \item The resulting colony will be quite big, and it can be difficult to image the full area or predict the specific spots where fruiting bodies will appear. Therefore, it is advised to pick at least a few frames of the original droplet perimeter, noticeable by the accumulation of cells, and a frame in the centre of the inoculation droplet. In such regions, fruiting bodies are the most likely to appear.
    
\end{enumerate}

\subsection{Predation of other organisms}

\begin{figure*}
 \centering
\includegraphics[trim=0 0 0 0 ,clip, scale=0.75
, angle=0,origin=c]{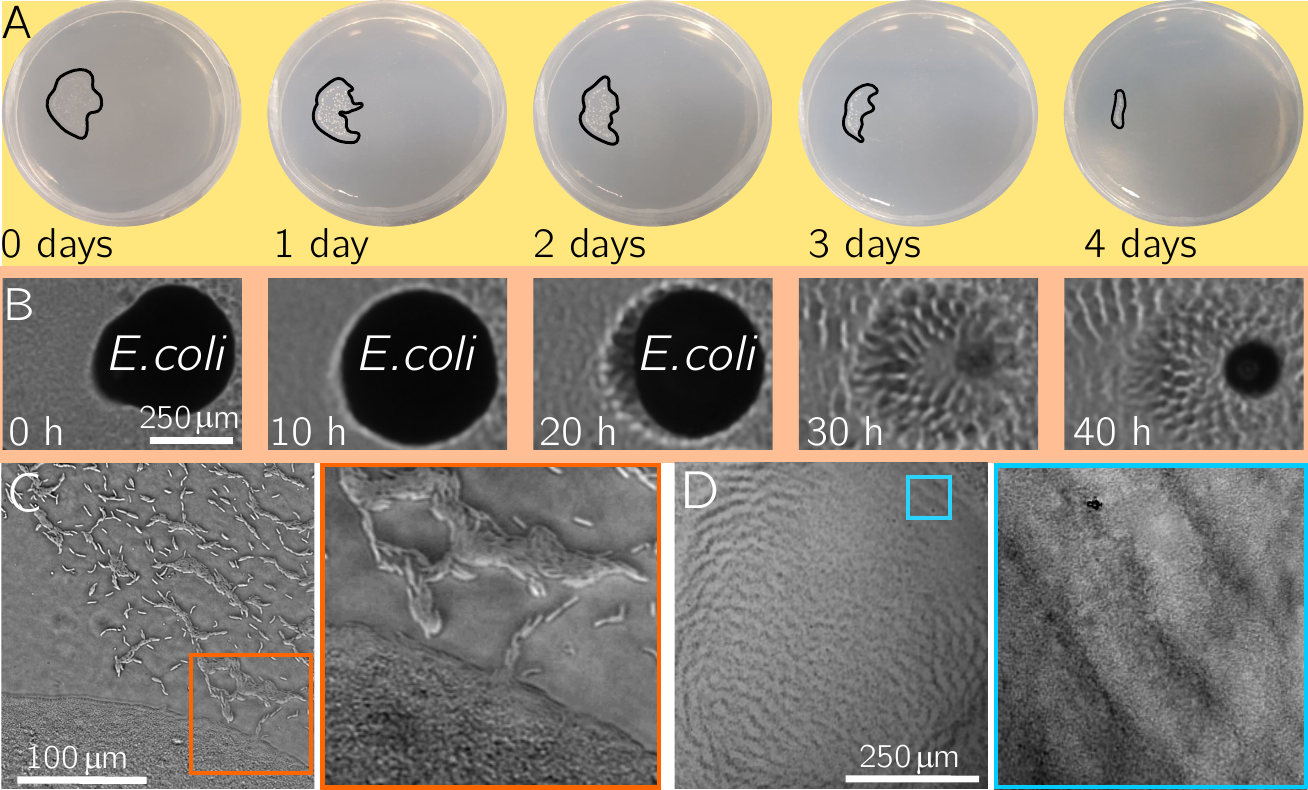}
\caption{{\bf Predation of \emph{Esterichia coli} by \textit{M. xanthus}.} 
{\bf A.} Timelapse of predation on \emph{E. coli} colonies (highlighted by a solid line) by \textit{M. xanthus} with images taken every day since predator/prey contact. The images show the full Petri dish of 55 mm of diameter over a black background to highlight the colonies. 
{\bf B.} Close-up on the progressive predation of a colony during a period of 40 hours since contact with a front of \textit{M. xanthus}. A fruiting body emerged at 40 hours. 
{\bf C.} Microscopic view of the first moment of predator/prey contact before rippling. {\bf D.} Fully assimilated colony at $10\times$ magnification, where rippling can be observed in the full surface of the colony with a close up at $63\times$ magnification. 
}
\label{fig:6}
\end{figure*}
    
\textit{M. xanthus} has been reported to feature predatory behaviour when it enters  contact with Gram negative bacteria\cite{Morgan2010,Thiery2020}, Gram positive bacteria\cite{Müller2016,Lloyd2017}, yeast\cite{Groß2023}, and other microorganisms. In this experiment, we use \emph{Escherichia coli} because it is the most widespread bacterial strain used in laboratories. However, using the proper medium and growth conditions, any other microorganism can be used in experiments instead. 

Even though predation can be observed in both TPM and CTT media, TPM is preferred to induce starvation and limit the source of nutrition exclusively to the prey\cite{Hillesland2007}. A colony of \textit{M. xanthus} in TPM is not perceptible, but the progressive shrinkage of the limits of prey domain, as shown in Figure~\ref{fig:6}A, can be easily seen by naked eye. 

In later stages, a waving pattern known as rippling will be formed by \textit{M. xanthus} on the prey colony\cite{Berleman2006}(see Supplementary Video 11). Rippling is the result of travelling waves of cells due to synchronised reversals in their trajectories\cite{Zhang2012,Bonilla2016}. Synchronisation of the full population improves their efficiency and represents a benefit for the whole community. The waves interact with each other at the edges of the colony forming an interference pattern as that shown in Figure~\ref{fig:6}B (see Supplementary Video 12). 

After the prey is fully assimilated, high density and starvation conditions are reached, leading to the potential formation of a fruiting body\cite{Berleman2007,Keane2016}.
At the cell level, \textit{M. xanthus} cells infiltrate and align along the edge of the colony before rippling starts, as shown in Figure~\ref{fig:6}C (see Supplementary Video 13). Rippling can still be appreciated at the microscopic scale; however the cell density involved to produce rippling phenomena does not allow to distinguish single cells in Figure~\ref{fig:6}D (see Supplementary Video 14), even though it is possible to replicate in a monolayer\cite{Sliusarenko2006}.

This is the protocol to realize these experiments:
\begin{enumerate}

    \item At the same time liquid cultures of \textit{M. xanthus} are prepared, inoculate liquid growth medium (lysogeny broth) inside a falcon tube with the strain of interest (\emph{E. coli}) in a clean environment.
    
    \item Grow overnight at the optimal temperature while shaking (37\degree{C} at 250 rpm).
    
    \item Prey cells are pelleted and washed in the same way as \textit{M. xanthus}. Put 1~mL of the cultivated media with prey in the desired concentration in an  Eppendorf tube. In this example, we use 0.5 $\text{OD}_{600}$.
    
    \item Centrifuge for 5 minutes at 8000 rpm ($4300g$ in relative centrifugal force).
    
    \item Open the lid of the Eppendorf tube in a clean area and get rid of the supernatant until left only with the pellet in the bottom of the tube.
    
    \item Add 1~mL of liquid TPM and close the lid.
    
    \item Resuspend with the vortex.
    
    \item Centrifuge for 5 minutes at 8000 rpm ($4300g$ in relative centrifugal force).
    
    \item Open the lid of the Eppendorf tube in a clean area and get rid of the supernatant until left only with the pellet in the bottom of the tube.
    
    \item Add $100\,{\rm \upmu L}$ (a 10th of the original quantity of the suspension) of liquid TPM. 
    
    \item The concentration of the prey can be varied as well according to the desired density of the target colony. The proposed concentration ensures a dense colony with no space between cells.
    
    \item Place a droplet of $13\,{\rm \upmu L}$ of \textit{M. xanthus} on the  surface of a TPM plate.
    
    \item Place a droplet of $5\,{\rm \upmu L}$ of the prey organism \emph{E. coli} on the surface of the agar plate, at a distance of approximately $0.5\,{\rm cm}$. The two droplets must not be in contact before they dry completely.
    
    \item Close the agar plate with the lid and seal the sides by using parafilm.
    \item Incubate at 32\degree{C} and observe frequently.
    \begin{enumerate}
    \item For macroscopic observation, take images every 1 hour for a total duration of 72 hours. 
    \item For microscopic observation, it is advised to start once both colonies are almost in contact, taking images every 2 to 10 minutes for at least 72 hours. 
    \item[] NOTE: The suggested times are only an orientation and development might still vary from one experiment to another.
    \end{enumerate}    
    
\end{enumerate}

\subsection{Preparation of substrates with microparticles}

\begin{figure*}
\centering
\includegraphics[trim=0 0 0 0 ,clip, scale=0.90
, angle=0,origin=c]{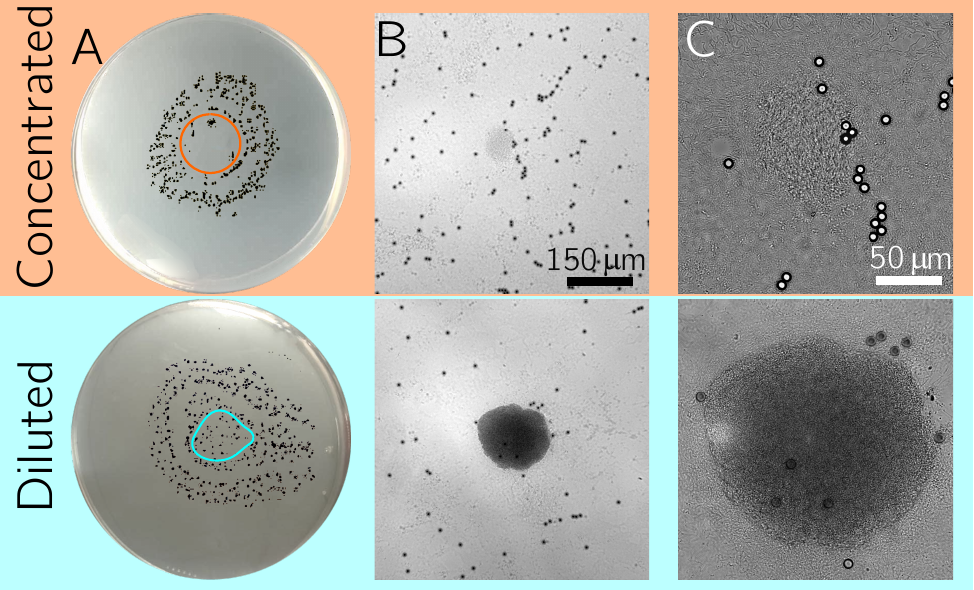}
\caption{{\bf Effect of microparticles on fruiting body formation.} 
{\bf A.} Comparison between substrates deposited with concentrated and diluted microparticles showing the full 55-mm-diameter Petri dish. The area where particles where deposited are circled for each case and the fruiting bodies highlighted in black.
{\bf B.} Comparison of both substrates at $20\times$ magnification showing a fruiting body.
{\bf C.} Close-up at $63\times$ magnification of each previous fruiting body where myxospores and cells can be distinguished.}
\label{fig:7}
\end{figure*} 

Microparticles introduce a level of complexity to the environment that can interfere with the displacement of cells and the emergence of collective behaviour. In the case of agar, the deposition of particles can strain the hydro-gel structure and in turn be detected by \textit{M. xanthus} as a prey colony (see Supplementary Video 15). The bacteria are then directed to the source of tension in the displacement surface in a behaviour known as elasticotaxis\cite{Stanier1942,Dworkin1983}. 

For this example, we use $7\,{\rm \upmu m}$  diameter melamine resin particles from  Microparticles Gmbh (MF-R-8060)
(note that different particle might produce different effects). In experiments, bacteria will be drawn to the particles and avoid further organisation\cite{Ramos2021}. As a consequence, a high concentration of microparticles  yields almost an absence of fruiting bodies in region where the particles were deposited to the naked eye in the same time frame, while a more diluted concentration yields more visible fruiting bodies in the particle deposited region as shown in Figure~\ref{fig:7}A. Still, fruiting bodies form normally in the particle-free space. Fruting bodies can still be observed in the vicinity of the microparticles, but the higher the concentration, the fewer and less developed these fruiting bodies will be, like in Figure~\ref{fig:7}B. At a microscopic level, less spores can be seen in the incipient fruiting body of concentrated substrate while diluted samples still show densely packed fruiting bodies after 96 hours as in Figure~\ref{fig:7}C. 

This is the protocol to realize these experiments:
\begin{enumerate}

    \item Suspensions of particles of approximately $7\,{\rm \upmu m}$ are prepared with different concentrations. In the case of this tutorial, the original microparticle solution of 10$\%$ (w/v) was further diluted to 1:500 and 1:2000, using Milli-Q water, resulting in a concentrated sample of $0.02\%$ (w/v) and a diluted one of 0.005$\%$ (w/v) respectively.
    
    \item Place a droplet of $50\,{\rm \upmu L}$ \  of the suspension of particles in the centre of a TPM plate and let it dry, preferably in a clean environment to avoid contamination.
    \begin{enumerate}
    \item If different types of particles need to be placed on the same agar plate, it is recommended to first turn over the plate and mark in the bottom the spots where each of the different droplets will be placed and label accordingly. Else, it is easy to lose track of the position of the particles, especially in low concentrations.
    \end{enumerate}
    
    \item Take pictures of different areas of the dried droplet and check that the number of particles is homogeneous, and that is consistent with the desired concentration. 
    \begin{enumerate}
    \item  To check the homogeneity, take pictures of different regions and count the relative particles.
    This step is particularly important to achieve reproducibility of the results.
    \item[] NOTE: If you are using different kinds of particles (for example melamine resin, polystyrene, or borosilicate), the initial concentrations will probably not be the same, so the concentrations will have to be adjusted accordingly.
    \end{enumerate}

    \item Once the protocol for the formation of substrates has been repeated in a controllable fashion, we can proceed to inoculate the bacteria.
    
    \item Place a droplet of $13\,{\rm \upmu m}$ \ of \textit{M. xanthus} in the centre of the droplet with particles.
    \begin{enumerate}
    \item For the macroscopic case, take images every hour for a total duration of 96 hours and compare the results between different concentrations of particles.
    \item For the microscopic case, take images every 10-20 minutes for a total duration of 96 hours. 
    \item[] NOTE: The suggested times are only an orientation and development might still vary from one experiment to another.
    \end{enumerate}
    
\end{enumerate}

\section{Conclusions}

This tutorial provides a comprehensive guide to culturing \textit{M. xanthus}, a bacterial strain with significant potential for active matter studies\cite{Cotter2017,AriasDelAngel2018,Zhang2018,Zhang2020,Balagam2021,Murphy2023}. While some studies have developed models to explain emergent patterns\cite{Sozinova2006, Starruß2007,Holmes2010,Balagam2015}, substantial knowledge gaps remain in capturing the multiscale behaviour of \textit{M. xanthus} and in validating these models against experimental data. Key areas for further exploration include transitions between behavioural stages, effects of external conditions, and studies in natural or nature-mimicking environments.

A major challenge in advancing this field has been the entry barrier for researchers without specific microbiological expertise. By lowering this barrier, we aim to foster interdisciplinary research and stimulate further investigation into the promising world of active matter and \textit{M. xanthus}. To this end, we present clear, comprehensive steps for culturing \textit{M. xanthus}, ensuring accessibility even for researchers new to bacterial culture. The reproducibility of these methods has been validated across different researchers and laboratories, reinforcing their reliability. Additionally, we describe simple yet informative experiments that can be performed with \textit{M. xanthus}, studying its behaviour at both micro- and macroscopic scales. These include observations of collective movement patterns and social behaviours, providing a solid starting point for more advanced studies.

Bridging the technical gap for researchers from various disciplines to work with \textit{M. xanthus}, this tutorial aims at catalysing innovative active matter research and at offering new insights into bacterial collective behaviours.

\section*{Author Contributions}

Jesus Manuel Antúnez Domínguez:
funding acquisition, investigation, methodology, data curation, software, visualization, writing – original draft.
\newline
Laura Pérez García: 
funding acquisition, investigation, methodology, data curation, visualization,  writing – original draft.
\newline
Natsuko Rivera-Yoshida:
funding acquisition, investigation, methodology, writing – review \& editing.
\newline 
Jasmin Di Franco:
investigation, validation, writing – review \& editing.
\newline 
David Steiner:
investigation, validation, writing – review \& editing.
\newline 
Alejandro V. Arzola:
funding acquisition, resources, supervision, writing – review \& editing.
\newline
Mariana Benítez:
funding acquisition, resources, supervision, writing – review \& editing.
\newline
Charlotte Hamngren Blomqvist:
resources, validation, writing – review \& editing.
\newline
Roberto Cerbino:
funding acquisition, resources, supervision, writing – review \& editing.
\newline
Caroline Beck Adiels:
conceptualization, funding acquisition, project administration, resources, supervision, writing – review \& editing.
\newline
Giovanni Volpe:
conceptualization, funding acquisition, project administration, resources, supervision, writing – review \& editing.

\section*{Conflicts of interest}

There are no conflicts of interest to declare.

\section*{Acknowledgements}

Jesus Manuel Antunez, Caroline Beck Adiels, and Giovanni Volpe acknowledge support from the MSCA-ITN-ETN project ActiveMatter sponsored by the European Commission (Horizon 2020, Project Number 812780).

Jesus Manuel Antunez and Laura Pérez García Dominguez thank the Adlerbertska forskningsstiftelsen program from years 2022 and 2023 respectively.

Natsuko Rivera, Caroline Beck Adiels, and Giovanni Volpe thank the Linnaeus-Palme International Exchange Program.

Natsuko Rivera-Yoshida, Mariana Benítez and Alejandro Vásquez gratefully acknowledge the financial support of the John Templeton Foundation (\#62220). The opinions expressed in this paper are those of the authors and not those of the John Templeton Foundation.

Giovanni Volpe acknowledge support from the project ERC-CoG project MAPEI sponsored by the European Commission (Horizon 2020, Project No. 101001267) and from the Knut and Alice Wallenberg Foundation (Grant No. 2019.0079).

\balance


\bibliography{rsc} 
\bibliographystyle{rsc}

\end{document}


\maketitle

 \begin{table}[h]
\centering
\small
  \caption{\ List of equipment used in the described protocol.}
  \label{tbl:equipment}
  \begin{tabular*}{1.20\textwidth}{ll}
    \hline
Equipment & Specifications  \\ \hline
Autoclave                                                                       & Front-loading benchtop autoclave RAYPA Model AH-21 B \\
Spectrophotometer                                                   &    Harvard Biochrom Ultrospec\textsuperscript{\texttrademark} 10 Cell Density Meter        \\ 
Orbital shaker & Fisherbrand\textsuperscript{\texttrademark} Digital Orbital Shaker (17811682)\\
Incubator                                                         &    Thermo Scientific\textsuperscript{\texttrademark} Heratherm\textsuperscript{\texttrademark} 
\\&General Protocol Microbiological Incubator (10529070)       \\ 
 Vortexer                                                                           &          \\
 Centrifuge                                                                      &   Eppendorf\textsuperscript{\texttrademark} Centrifuge 5430 R - \\&Refrigerated Microcentrifuge           \\ 
-80\degree{C} freezer                                                &  Thermoscientific ULT freezer (TDE30086FV5ID)\\
Water Purification system                                       &  ELGA PURELAB\textsuperscript{\textregistered} Chorus 1 Complete  \\ 
 Scale                                                                                &    Kern \& Sohn Precision balance EG 600-3AM      \\
 Pipette 100-1000 $\mu$l      &   LABSOLUTE\textsuperscript{\textregistered}  Microliter pipette 100-1000 $\mu$l\\ 
 pH-meter                                                                        &    Fisherbrand\textsuperscript{\texttrademark} accumet\textsuperscript{\textregistered} \\
                                                                                              & AB150 pH Benchtop Meters      \\
 Inoculating loop                                                           &   Microspec Sterile Plastic Inoculation Loops (Volume: 5$\mu$L) \\ 
 PCR UV cabinet                                                                               &    Grant Instrument UVT-B-AR  \\
 $70 \%$ Ethanol                        &   Contec Filtered Alcohols: 70\% IPA   \\
500~mL Screw-top glass bottle                                       &  Fisherbrand\textsuperscript{\texttrademark}  Reusable Glass Media Bottles with Cap (15476113)   \\
100~mL Erlenmeyer glass flask                              &  LABSOLUTE\textsuperscript{\textregistered} Th Geyer (7690027)  \\
 \hline
\end{tabular*}
\end{table}